\documentclass[aps,pra,twocolumn,superscriptaddress]{revtex4}%
\usepackage{amssymb}
\usepackage{graphicx}
\usepackage{amsmath}
\usepackage{amsbsy}
\usepackage{amsfonts}
\usepackage{amssymb,dsfont,physics}
\usepackage{tikz}
\usepackage{accents}%
\providecommand{\U}[1]{\protect\rule{.1in}{.1in}}
\begin{document}
\title{Quantum channel position finding using single photons}
\author{Athena Karsa}
\affiliation{Department of Computer Science, University of York, York YO10 5GH, UK}
\author{Jacques Carolan}
\affiliation{Wolfson Institute for Biomedical Research, University College London, Gower Street, London WC1E 6BT, UK}
\author{Stefano Pirandola}
\affiliation{Department of Computer Science, University of York, York YO10 5GH, UK}
\date{\today}

\begin{abstract}
Channel position finding is the task of determining the location of a single target channel amongst an ensemble of background channels. It has many potential applications, including quantum sensing, quantum reading and quantum spectroscopy. In particular, it could allow for simple detection protocols to be extended to ones of measurement, for example, target ranging with quantum illumination. The use of quantum states and entanglement in such protocols have shown to yield quantum advantages over their optimal classical counterparts. Here we consider quantum channel position finding using sources specified by at most one single photon on average per mode, using the discrete variable formalism. By considering various quantum sources it is shown through the derivation of performance bounds that a quantum enhancement may be realised.
\end{abstract}

\maketitle

\section{Introduction}

Quantum hypothesis testing (QHT)~\cite{Helstrom,chefles2000quantum} is a fundamental tool of quantum sensing where the goal is to distinguish between two alternate hypotheses with applications in quantum reading~\cite{pirandola2011reading,pirandola2018advances}, quantum illumination~\cite{tan2008quantum,karsa2020generic} and spectroscopy~\cite{mukamel2020roadmap}. These are problems of quantum channel discrimination (QCD)~\cite{sacchi2005entanglement} whereby the different scenarios to be distinguished are characterised by different physical processes modelled as quantum channels of varying parameters. By employing sources which are sensitive to these parameters as channel inputs, the outputs may be analysed in order to make a decision. Determining the optimal strategy for such a task becomes an optimisation problem over both input states and output measurements, typically subject to energetic constraints on the total number of probings and, in the bosonic case, mean number of photons employed.

Channel position finding (CPF)~\cite{zhuang2020entanglement,pereira2021idler} is a problem in multiple channel discrimination whereby given an array of quantum channels where all but one is different, the task is to locate the dissimilar one. It allows for QHT to be extended beyond well-studied binary problems. In the case of quantum illumination, one may be certain that the target is located within a region of interest but is interested in determining precisely where, as in quantum target finding~\cite{zhuang2020entanglement} (on a plane) or quantum target ranging~\cite{karsa2020energetic}. The protocol for such a task would involve probing each of the pre-defined locations a fixed number of times then collecting and performing suitable measurements on the returning states. Within the CPF framework, the differing pathways in terms of loss and target reflectivity may be encoded within the quantum channels under study.

Zhuang and Pirandola~\cite{zhuang2020entanglement} formulated the general problem of CPF for the testing of multiple quantum hypotheses providing upper- and lower-bounds on the error probability. This was given for a classical CPF protocol using coherent state sources (minimum uncertainty states with positive P-representation, considered classically `optimal') and compared to a specific quantum protocol employing maximally entangled two-mode squeezed vacuum (TMSV) states. It was shown that by using the generalised conditional-nulling (CN) receiver at the output a quantum advantage could be achieved.

In this paper, we study the problem of CPF subject to the constraint that the sources considered are comprised of \emph{at most} one single photon. Remaining in the discrete-variable setting, while works so far have been focused on continuous-variables, we study the potential of quantum-enhanced CPF for various source specifications: Single-photon (Fock) state, GHZ state exhibiting multipartite entanglement across the channel array, bipartite (signal-idler) entanglement in a quantum illumination-style setup, and biphoton (Bell) states. In Section~\ref{sec:probspec} we outline the problem of CPF describing the model and mathematical tools used to quantify performances with respect to the source. In Section~\ref{sec:results} we consider the four quantum sources previously described providing formulae for fidelity and bounds on the CPF error probability. These are compared to the classical benchmark of a coherent state whose amplitude is $\alpha =1$. Finally, in Section~\ref{sec:receiver} a simple receiver based on photon-counting is outlined and applied to two cases: classical coherent state source and single-photon (Fock) state source.

\section{Problem Specification}\label{sec:probspec}

\subsection{Basic model}

\begin{figure}[t]
\centering
\includegraphics[width=0.9\linewidth]{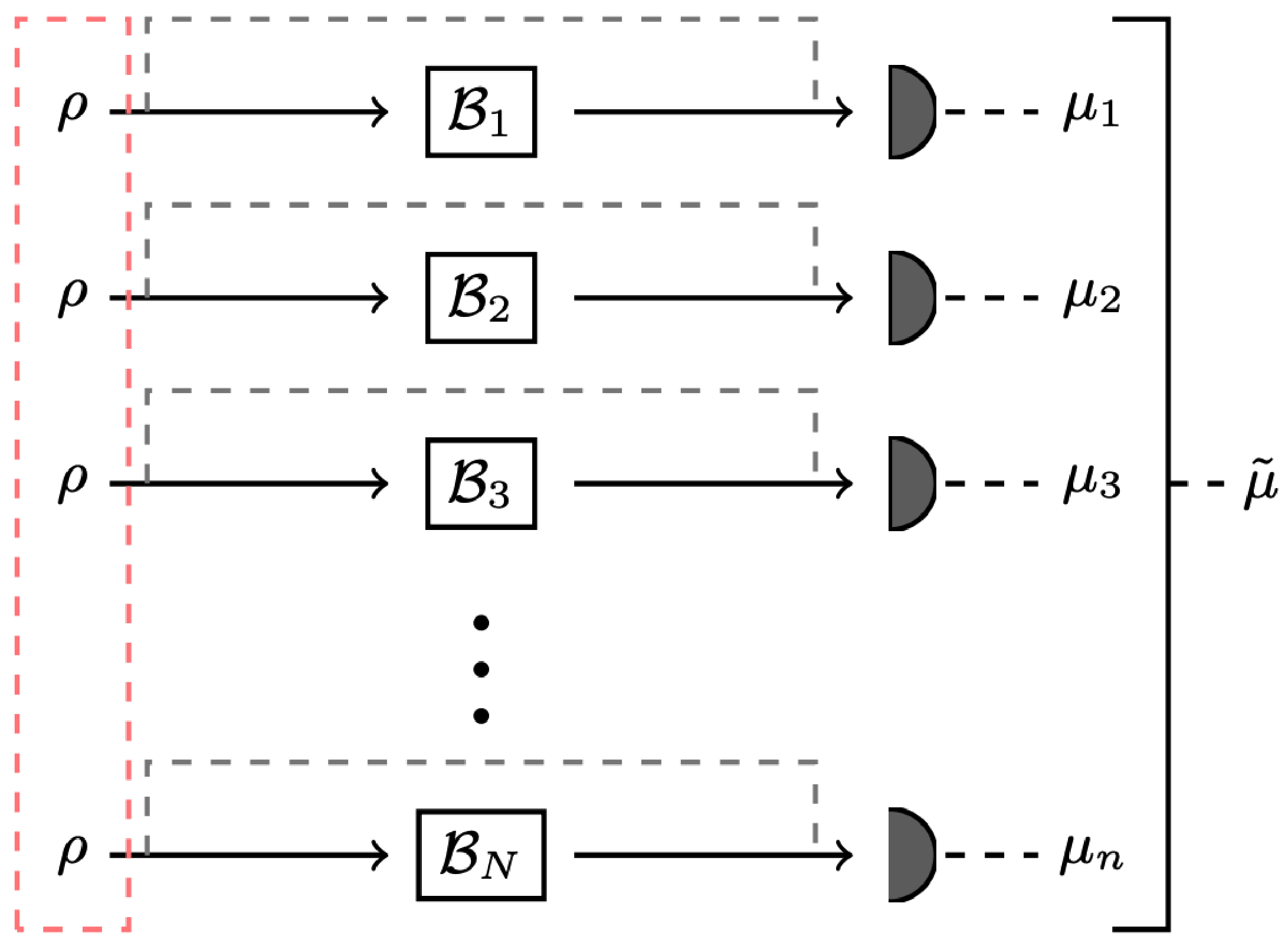}
\caption{Schematic diagram of channel position finding (CPF). For $i=1,\dots, N$ we have an ensemble of boxes $B_i$ and the task is locate the single target channel amongst background channels using copies of the source, $\rho$, which may be specified in several ways. In the classical (coherent state) and single-photon strategies, the source is sent through the channel (black) with the output going straight to the receiver for post-processing. In the quantum strategy two scenarios exist: first, a maximally-entangled two-mode source comprising a signal (black), sent through the channel, and an idler (grey) which recombines with the output at the receiver; second, there are no idlers however the source comprises of multi-mode entanglement spanning the entire $N$-box array (red). In each case, outputs from boxes are combined to yield a final result, $\tilde{\mu}$, giving the target's location with some error probability.}
\label{diagram}
\end{figure}

Consider the basic model of quantum CPF comprised of $N \geq 2$ input-output black boxes as shown in Fig.~\ref{diagram}. For $i=1,2,\dots N$, the $i$th box $\mathcal{B}_i$ contains either a reference channel $\mathcal{R}$ or some target channel $\mathcal{E} \neq \mathcal{R}$ and the task is to locate $\mathcal{E}$. We assume the target channel only occupies one box, such that joint probabilities of the form $\mathrm{P}(\mathcal{B}_i=\mathcal{E},\mathcal{B}_j = \mathcal{E})$ are all zero, and the target channel is in one of the boxes with certainty, i.e., $\mathrm{P}(\mathcal{B}_i =\mathcal{R}\, \forall i)=0$. Identification of the target channel is a problem of symmetric quantum hypothesis testing where the task is to discriminate between $N$ hypotheses given by
\begin{equation}
H_i : \mathcal{B}_i = \mathcal{E} \quad \mathrm{and} \quad \mathcal{B}_{j \neq i} = \mathcal{R}.
\end{equation}

To carry out this task we employ a quantum state $\rho$ as input into each of the boxes $\mathcal{B}_i$. Such a quantum system interacting with its environment unavoidably undergoes the quantum dynamical process of decoherence. Such a noisy process may be described by the CPT map $\mathcal{B}(\rho)$ acting on the quantum state $\rho$ using the Kraus representation
\begin{equation}
\mathcal{B}(\rho) = \sum_{\mu} K_{\mu} \rho K_{\mu}^{\dag},
\end{equation}
where $K_{\mu}$ are the Kraus operators satisfying $\sum_{\mu} K_{\mu}^{\dag}K_{\mu}=\mathds{1}$. The output state is then clearly dependent on the channel's specification: whether $\mathcal{B} =\mathcal{R}$, the reference channel, or $\mathcal{B} =\mathcal{E}$, the target channel.

We model the contents of each box $\mathcal{B}_i$ as a purely dissipative amplitude damping channel (ADC) with Kraus operators~\cite{nielsen2002quantum}
\begin{equation}
\mathbf{K}_{\mathrm{ADC}}= \{ \dyad{0}{0} + \sqrt{1-\gamma}\dyad{1}{1}, \sqrt{\gamma} \dyad{0}{1} \},
\end{equation}
where $\gamma$ is the damping rate. Of course, in the case where $\gamma=0$ the channel reduces to the identity. Such a model allows us to also consider inefficient detectors which may themselves be modelled as ADCs with damping rate $\gamma=1-\eta$, where $\eta$ is the efficiency.

The classical output at each detector takes a binary value $\mu_i \in \{0,1\}$ corresponding to a decision on whether $\mathcal{B}_i =\mathcal{R}$ or $\mathcal{B}_i =\mathcal{E}$, respectively. All of these $N$ outputs must then be post-processed to give a final decision $\tilde{\mu}$ on the target channel's position.

\subsection{Quantum state discrimination}

The task of quantum CPF may be reduced to one of quantum state discrimination~\cite{Helstrom,chefles2000quantum}. Our aim is, for a given input state $\rho$, to best distinguish between two possible channel outputs $\mathcal{E}(\rho)$ and $\mathcal{R}(\rho)$.

To determine whether or not a quantum advantage exists we must compute and compare the error probabilities $p_{\mathrm{err}}^{N,M} (\rho)$ for the $N$-box, $M$-use discrimination problem using a classical state $\rho_{\mathrm{C}}$ and quantum state $\rho_{\mathrm{Q}}$ as input. Exact analytical forms of error probability are difficult to compute but may be replaced by upper and lower bounds~\cite{bagan2016relations,ogawa1999strong} such that for any input state $\rho$ we may write
\begin{equation}
L(\rho) \leq p_{\mathrm{err}}^{N,M} (\rho) \leq U(\rho).
\end{equation}
Then, establishing that the inequality $U(\rho_{\mathrm{Q}}) < L(\rho_{\mathrm{C}})$ holds is sufficient to prove that $p_{\mathrm{err}}^{N,M}(\rho_{\mathrm{Q}}) < p_{\mathrm{err}}^{N,M}(\rho_{\mathrm{C}})$.

We assume equiprobable hypotheses, so that $p_i=N^{-1}$ for any $i$. Suppose that the overall input state has a tensor product form over the $N$ boxes such that $\rho = \sigma ^{\otimes N}$, such that
\begin{equation}
\mathcal{E}_i^N (\sigma^{\otimes N}) = \bigotimes_{j \neq i} \mathcal{R}_j(\sigma) \otimes \mathcal{E}_i(\sigma).
\label{tensorinput}
\end{equation}
Of course, Eq. (\ref{tensorinput}) does not hold for more elaborate quantum systems such as GHZ states where multipartite entanglement is distributed across the entire arrangement of boxes (this will be studied in Section \ref{multipartitesec}). Then, we have the following upper- and lower-bounds~\cite{barnum2002reversing,zhang2001upper,montanaro2008lower,qiu2010minimum}, respectively, for the error probability
\begin{equation}\label{eq:UB}
p_{\mathrm{err}}^{N,M} (\rho) \leq (N-1) F^{2M} \left( \mathcal{E}(\sigma), \mathcal{R}(\sigma) \right),
\end{equation}
\begin{equation}\label{eq:LB}
p_{\mathrm{err}}^{N,M} (\rho) \geq \frac{N-1}{2N}  F^{4M} \left( \mathcal{E}(\sigma), \mathcal{R}(\sigma) \right),
\end{equation}
where $F$ is the Bures' fidelity~\cite{jozsa1994fidelity,uhlmann1976transition} 
\begin{equation}\label{eq:FID}
F(\rho ,\sigma):= || \sqrt{\rho} \sqrt{\sigma}||_1 = \Tr \sqrt{\sqrt{\rho} \sigma \sqrt{\rho}}.
\end{equation}

\section{Results}\label{sec:results}

In the following, Eqs.~(\ref{eq:UB}), (\ref{eq:LB}) and (\ref{eq:FID}) are used to compute the fidelity-based upper- and lower-bounds in error probability for the $N$-channel, $M$-use CPF protocol for the various sources under consideration. Namely, and in order of consideration, these are: the classical benchmark using coherent states; the single-photon (Fock) state; the GHZ state with multipartite entanglement; a two-mode (signal-idler) state with bipartite entanglement. For the final source type with bipartite entanglement an alternative, idler-free CPF protocol is provided. In all cases the final performance bounds are computed and plotted in Figs.~\ref{PerrBoundsIdentity} and~\ref{PerrBoundsDelta} at the end of Section~\ref{SECbipartite}.

\subsection{Classical benchmark}\label{classicalbenchmark}

For our classical benchmark we consider the classical input state $\rho_{\mathrm{C}} = \sigma_{\mathrm{C}}^{\otimes N}$ with $\sigma_{\mathrm{C}}= \dyad{\alpha}{\alpha}$. This is  a coherent state with Fock basis representation~\cite{RMP}
\begin{equation}
\ket{\alpha} = \exp \left( \frac{-|\alpha|^2}{2} \right) \sum_{n=0}^{\infty} \frac{\alpha^n}{\sqrt{n !}} \ket{n},
\end{equation}
with complex amplitude $\alpha = |\alpha|^2 e^{i \theta}$ where the magnitude $|\alpha|^2 = \bar{n}$ is the mean number of photons and $\theta$ is the phase. Under the action of an arbitrary ADC with damping rate $\gamma$, coherent state $\ket{\alpha} \rightarrow \ket{\sqrt{\tau} \alpha}$, where $\tau=1-\gamma$ is the transmissivity.

The squared-fidelity between the outputs of two ADCs, parametrised by rates $\gamma_0$ and $\gamma_1$, respectively, acting on $\ket{\alpha}$ is given by 
\begin{equation}
F^2 \left( \ket{\sqrt{\tau_0} \alpha}, \ket{\sqrt{\tau_1} \alpha} \right) = \exp \left(-\bar{n} ( \sqrt{1-\gamma_0} - \sqrt{1-\gamma_1})^2 \right),
\end{equation}
yielding the classical lower-bound benchmark
\begin{equation}
p_{\mathrm{err}}^{N,M} (\rho_{\mathrm{C}}) \geq \frac{N-1}{2N} \exp \left( - 2M ( \sqrt{1-\gamma_0} - \sqrt{1-\gamma_1})^2\right),
\label{CSlowerbound}
\end{equation}
where we have set $\bar{n}=1$.

\subsection{Single-photon state}\label{singlephoton}

The first quantum source we will consider is the tensor product of single-qubit or single-photon states $\rho_{\mathrm{Q}} = \sigma_{\mathrm{Q}}^{\otimes N}$ with $\sigma_{\mathrm{Q}}= \ket{1}$ in the computational basis. Under action of the ADC we have that
\begin{equation}
\sigma_{\mathrm{Q}} \rightarrow \gamma \dyad{0}{0} + (1-\gamma)\dyad{1}{1}.
\end{equation}
Computing the fidelity between two arbitrary ADC outputs we obtain
\begin{equation}
F\left( \rho_{\mathrm{Q},0}, \rho_{\mathrm{Q},1} \right) = \sqrt{(1-\gamma_0)(1-\gamma_1)}+ \sqrt{\gamma_0 \gamma_1},
\end{equation}
allowing us to write both the lower- and upper-bounds for error probability as
\begin{equation}
p_{\mathrm{err}}^{N,M} \left( \rho_{\mathrm{Q}} \right) \geq \frac{N-1}{2N} \left(  \sqrt{(1-\gamma_0)(1-\gamma_1)}+ \sqrt{\gamma_0 \gamma_1} \right)^{4M},
\label{Focklowerbound}
\end{equation}
and
\begin{equation}
p_{\mathrm{err}}^{N,M} \left( \rho_{\mathrm{Q}} \right) \leq (N-1) \left(  \sqrt{(1-\gamma_0)(1-\gamma_1)}+ \sqrt{\gamma_0 \gamma_1} \right)^{2M}.
\label{Fockupperbound}
\end{equation}

\subsection{Multipartite entanglement}\label{multipartitesec}

Of course one may consider a quantum source with entanglement distributed across the $N$ boxes. Such a source could take the form of a GHZ state, an entangled quantum state of $N>2$ $d$-dimensional subsystems, given by
\begin{equation}
\ket{\mathrm{GHZ}} = \frac{1}{\sqrt{d}} \sum_{i=0}^{d-1} \ket{i}^{\otimes N}.
\end{equation}
In the case of qubits ($d=2$), it reads
\begin{equation}
\ket{\mathrm{GHZ}} = \frac{\ket{0}^{\otimes N} + \ket{1}^{\otimes N} }{\sqrt{2}}.
\end{equation}
Consider now the action of an $N$-box system consisting of ADCs with damping rate $\gamma_0$ or $\gamma_1$ (for one of the $N$ boxes) on the $N$-partite GHZ state $\Psi = \dyad{\mathrm{GHZ}}{\mathrm{GHZ}}$. The resulting output state consists of all possible partial decays of constituent states $\ket{1} \rightarrow  \ket{0}$. Clearly, such a state does not have a tensor product form across the $N$ boxes so the fidelity must be computed across the entire $N$-partite system outputs
\begin{equation}
\mathcal{E}_i^N (\Psi) = \bigotimes_{j\neq i} \mathcal{R}_j \otimes \mathcal{E}_i (\Psi),
\end{equation}
such that the error probability upper bound reads
\begin{equation}
p_{\mathrm{err}}^{N,M} (\rho) \leq (N-1) F^{M} \left( \mathcal{E}_i^N(\Psi), \mathcal{E}_k^N(\Psi) \right).
\label{GHZupperbound}
\end{equation}

\subsection{Bipartite (signal-idler) entanglement}\label{SECbipartite}

\begin{figure}[t]
\centering
\includegraphics[width=.9\linewidth]{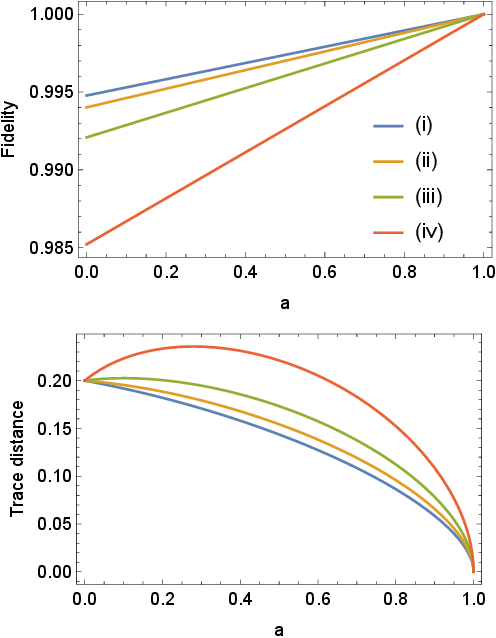}
\caption{Plots showing fidelity (upper) and trace distance (lower) as a function of $a$ for varying values of damping rate $\gamma_0=\gamma_1+0.1$ with (i) $\gamma_1=0.55$, (ii) $\gamma_1=0.65$, (iii) $\gamma_1=0.75$, and (iv) $\gamma_1=0.85$. Lines appear in this order, top to bottom, in the upper panel while the ordering is reversed in the lower panel. Fidelity is always minimised for single qubit/photon states where $a=0$. The trace distance is maximised in most cases by such a Fock state however the optimum value of $a$ resides somewhere in between the two extremes considered, i.e., $0 < a < 1/2$ in regions where damping rates are both relatively high ($\gtrsim 0.75$).}
\label{GenStateFidTD}
\end{figure}

An alternative, entanglement-based quantum source is given by a tensor product over all the boxes ($\otimes N$) where each signal $S_i$ is coupled to an ancillary idler $I_i$. Only the signal probes the box while the idler is sent directly to the receiver to join the output. The total joint-state $\Phi_{S,I}$ forms a Bell pair in the case where the two modes are maximally-entangled and the associated quantum channel takes the form
\begin{equation}
\mathcal{E}_i^{N}:= \otimes_{j \neq i} (\mathcal{R}_{S_j} \otimes \mathcal{I}_{I_j}) \otimes (\mathcal{E}_{S_i} \otimes \mathcal{I}_{I_i}).
\label{bipartitechannel}
\end{equation}
Consider as our two-mode source $\Phi_{S,I}$ a generic state with density operator given by
\begin{equation}\label{rhogen}
\begin{split}
\rho_{\mathrm{gen}} =  a \dyad{00}{00} &+ \sqrt{a (1-a)} \left( \dyad{00}{11} + \dyad{11}{00}  \right) \\
&+ (1-a)\dyad{11}{11},
\end{split}
\end{equation}
where $a$ is a parameter quantifying how close $\rho_{\mathrm{gen}}$ is to a Bell state. That is, when $a = 1/2$ the source is a maximally-entangled Bell state while when $a=0$ (or $a=1$) we have the single qubit/single photon state $\ket{11}$ (or vacuum state $\ket{00}$), which is clearly separable.

Upon action of an ADC with damping rate $\gamma$ only on the signal ($S$) mode while performing the identity on the reference idler ($I$) mode our output joint state reads
\begin{equation}
\begin{split}
\rho_{\mathrm{gen}} \rightarrow \rho_{\mathrm{gen}}'=  &a \dyad{00}{00} + (1-a)\gamma \dyad{01}{01} \\
&+ \sqrt{a (1-a) (1-\gamma)} (\dyad{00}{11} + \dyad{11}{00} \\
&+ (1-a)(1-\gamma)\dyad{11}{11}.
\end{split}
\end{equation}
Computing the fidelity between two output states under differing ADCs with rates $\gamma_0$ and $\gamma_1$ we obtain
\begin{equation}\label{plottedfid}
\begin{split}
F\left(\rho_{\mathrm{gen},0}' , \rho_{\mathrm{gen},1}' \right) = (1-a)&\sqrt{(1-x^2)(1-y^2)}\\
&+ a + xy - axy,
\end{split}
\end{equation}
where we have defined $x=\sqrt{1-\gamma_0}$ and $y=\sqrt{1-\gamma_1}$.

It is clear that the state $\rho_{\mathrm{gen}}$ achieving a minimal discrimination error corresponds to that also minimising the above fidelity. Performing this minimisation over $a$ gives the minimum fidelity:
\begin{equation}
F_{\mathrm{min}}\left( \rho_{\mathrm{gen}}' \right) = \begin{cases}
1 & x=y\\
x & x<y=1\\
y & y<x=1\\
xy + \sqrt{(x^2-1)(y^2-1)} & \mathrm{otherwise}.
\end{cases}
\end{equation}
These minima are achieved for $a=0$ corresponding to the single qubit/photon Fock state $\ket{1}$. The trivial solution of $F_{\mathrm{min}}(\rho_{\mathrm{gen}}') = 1$ for all values of $x=y$ is achieved for both $a=0,1/2$. 

The upper panel of Fig.~\ref{GenStateFidTD} plots the quantum fidelity given in Eq.~(\ref{plottedfid}) as a function of $a$ for varying values of $\gamma_0$ and $\gamma_1$, confirming that the minimisation occurs at $a=0$, corresponding to a Fock state. Thus, for use in a signal-idler set-up CPF protocol, Bell states are sub-optimal sources compared to Fock states. Another tool for quantifying the distinguishability of quantum states is the trace distance, where the minimum error may be achieved through its maximisation. This function is similarly plotted in the lower panel of Fig.~\ref{GenStateFidTD}.

\subsection{Biphoton states via integrate quantum photonics}

\begin{figure}[t]
\centering
\includegraphics[width=0.9\linewidth]{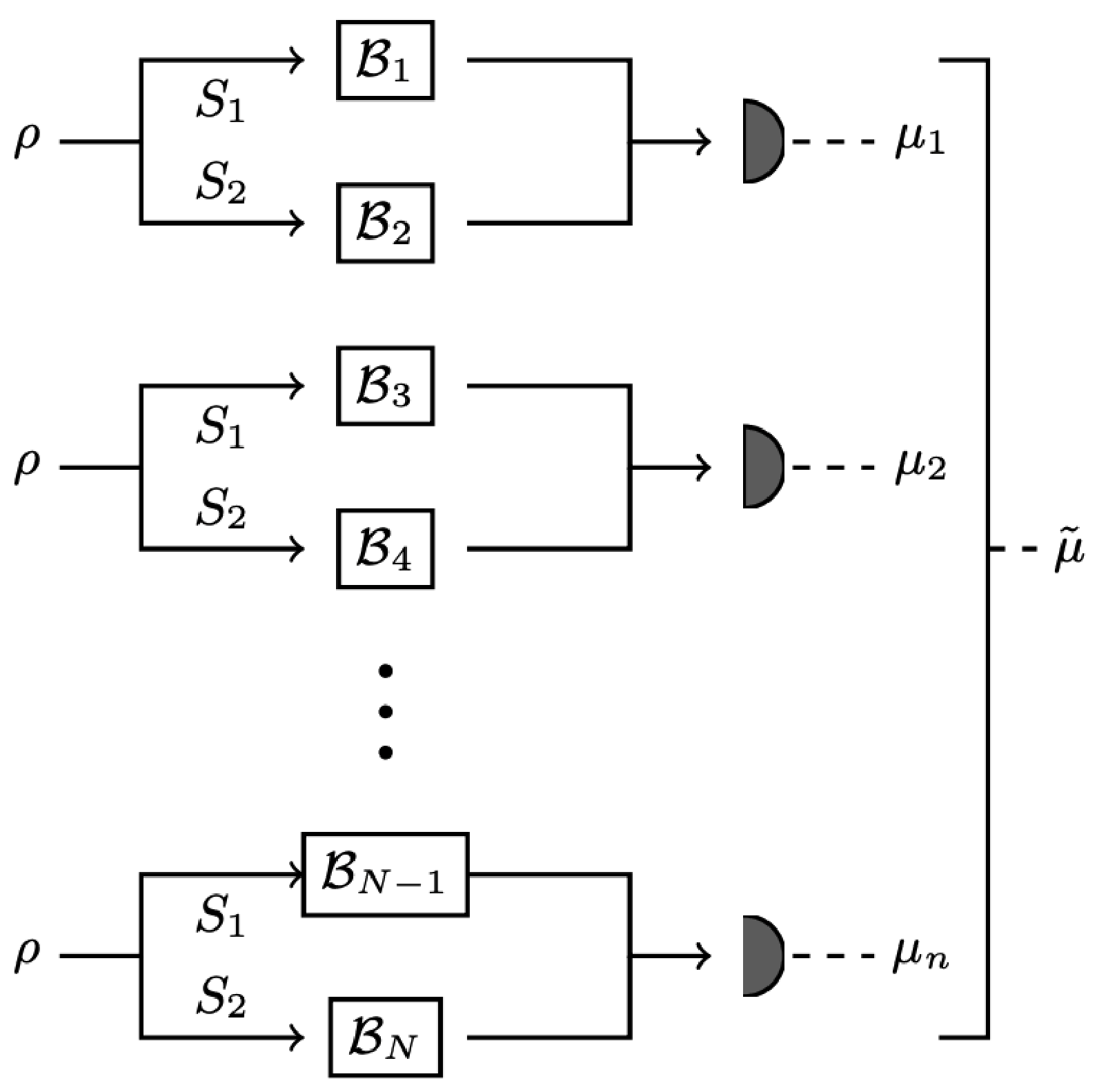}
\caption{Schematic diagram of idler-free channel position finding (CPF) for biphoton states exhibiting bipartite entanglement for channel arrays consisting of even $N$ boxes.}
\label{diagrambiphoton}
\end{figure}

A leading approach to generating bright and pure single photon states on chip is through the use of microring resonators (MRRs)~\cite{clemmen2009continuous,silverstone2015qubit,caspani2017integrated}. They comprise a waveguide ring coupled to a bus waveguide that produces well-defined resonances when the ring's circumference is an integer number of wavelengths~\cite{bogaerts2012silicon}. Further, when the waveguide itself is made of a $\chi^{(3)}$ material, such as silicon~\cite{silverstone2014onchip} or silicon nitride~\cite{dutt2015onchip}, photons can be generated via the spontaneous four-wave mixing process. Specifically, pumping the ring with a bright laser at frequency $\omega_p$ can cause two photons to be absorbed from the pump, generating a pair of photons at neighbouring frequencies $\omega_s$ and $\omega_i$ such that energy and momentum are conserved. The frequencies of these generated photons are thus $\omega_s = \omega_p + i\times \mathrm{FSR}$ and $\omega_i = \omega_p - i\times \mathrm{FSR}$ where FSR is the free spectral range of the rings and $i$ indexes a particular resonance. This state is typically referred to as a `biphoton' state and, assuming the weak pumping regime, can be written as
\begin{equation}
    \ket{\psi} = \frac{1}{\sqrt{n}} \sum_{i=1}^n \ket{\omega_{-i}} \ket{\omega_{+i}}.
    \label{biphotonstate}
\end{equation}
Here $\ket{\omega_i}$ represents a single photon in the $i$th frequency mode and $n$ typically depends on phase matching conditions for the MRR, which can reach up to $n=40$ modes~\cite{imany2018microresonator}. 

\subsubsection{Use in a signal-idler protocol}

When $n=2$ the output is a Bell state in the frequency basis given by
\begin{equation}
    \ket{\psi_2} = \frac{1}{\sqrt{2}}(\ket{\omega_{-1}}\ket{\omega_{+1}} + \ket{\omega_{-2}}\ket{\omega_{+2}}). 
    \label{biphotonbell}
\end{equation}
We have seen in Sec. \ref{SECbipartite} that such states, used in a signal-idler type set-up for each box, are sub-optimal compared to pure single-photon states. They are, however, relatively straightforward to generate on chip at both visible and NIR frequencies, and can have a frequency resolution of $\sim 1$ pm (100's MHz)~\cite{ramelow2015siliconnitride}.

Consider the biphoton Bell state (Eq.~(\ref{biphotonbell})) in a signal-idler set-up for the task of quantum channel position finding, subject to the quantum channel given by Eq.~(\ref{bipartitechannel}). In such a protocol we send the first, $-i$, signal mode $S_i$ into the box while retaining the second, $+i$, idler mode $I_i$ for later joint measurement. Computing the fidelity between the two two-mode output states under ADCs with rates $\gamma_0$ and $\gamma_1$ we obtain
\begin{equation}
\begin{split}
    F(\dyad{\psi_2}{\psi_2}_0,&\dyad{\psi_2}{\psi_2}_1) =\\
    &\frac{1}{2}\left( \sqrt{1-\gamma_0} \sqrt{1-\gamma_1} + 1 + \sqrt{\gamma_0 \gamma_1}\right),
\end{split}
\end{equation}
yielding the lower- and upper-bounds
\begin{equation}
     p_{\mathrm{err}}^{N,M}(\dyad{\psi_2}{\psi_2}) \geq \frac{N-1}{2N} F(\dyad{\psi_2}{\psi_2}_0,\dyad{\psi_2}{\psi_2}_1)^{4M},
\end{equation}
and
\begin{equation}
    p_{\mathrm{err}}^{N,M}(\dyad{\psi_2}{\psi_2}) \leq (N-1) F(\dyad{\psi_2}{\psi_2}_0,\dyad{\psi_2}{\psi_2}_1)^{2M}.
\end{equation}

\subsubsection{Use in an idler-free protocol}

A CPF protocol may be devised in which the full entanglement exhibited in a Bell state may be exploited across the multi-channel array in order to realise a quantum advantage. Such an advantage may be readily demonstrated due to the experimental availability of biphoton states. 

Consider the CPF protocol for an even number of boxes $N$. Using as a source the two-mode biphoton state of Eq.~(\ref{biphotonbell}), label each mode as a signal, $S_1$ and $S_2$, to be used as a probe between two adjacent boxes, as shown in Fig.~\ref{diagrambiphoton}. Then for any CPF problem comprising $N \geq 4$ individual channels, the global quantum channel acting on the state is 
\begin{equation}
\mathcal{E}_i^{N/2}:= \otimes_{j \neq i} (\mathcal{R}_{S_{1,j}} \otimes \mathcal{R}_{S_{2,j}}) \otimes (\mathcal{E}_{S_{1,i}} \otimes \mathcal{R}_{S_{2,i}}).
\label{idlerfreechannel}
\end{equation}
Computing the fidelity between the two two-mode output states under ADC pairs with rates ($\gamma_0,\gamma_0$) and ($\gamma_1,\gamma_0$) we obtain
\begin{equation}
    F(\dyad{\psi_2}{\psi_2}_{0,0},\dyad{\psi_2}{\psi_2}_{1,0}) =\frac{1}{2\sqrt{2}}\sqrt{\frac{\alpha}{ \beta}}
,
\end{equation}
where 
\begin{equation}
  \alpha =2 + 4 \Delta + 2 \Gamma + 4 \Delta \Gamma + 4 \Delta \gamma_0 - \gamma_1 + \gamma_0(1 + 2\Gamma + \gamma_0 + 5 \gamma_1 )
,
\end{equation}
\begin{equation}
  \beta = (1 + \Delta)(1 + \gamma_0)
,
\end{equation}
and we have defined
\begin{equation}
\begin{split}
    \Gamma &= \sqrt{1 - \gamma_0} \sqrt{1 - \gamma_1},\\
    \Delta &= \sqrt{\gamma_0 \gamma_1}.
\end{split}
\end{equation}
Then, the even $N$ idler-free CPF protocol has the following lower and upper bounds on the error probability in identifying the correct pair of channels of which one is the target:
\begin{equation}\label{idlerfreeerr1}
     \tilde{p}_{\mathrm{err}}^{N,M}(\dyad{\psi_2}{\psi_2}) \geq \frac{N-2}{2N} F(\dyad{\psi_2}{\psi_2}_{0,0},\dyad{\psi_2}{\psi_2}_{1,0})^{4M},
\end{equation}
and
\begin{equation}\label{idlerfreeerr2}
    \tilde{p}_{\mathrm{err}}^{N,M}(\dyad{\psi_2}{\psi_2}) \leq \frac{N-2}{2} F(\dyad{\psi_2}{\psi_2}_{0,0},\dyad{\psi_2}{\psi_2}_{1,0})^{2M}.
\end{equation}

\begin{figure}[t]
\centering
\includegraphics[width=.9\linewidth]{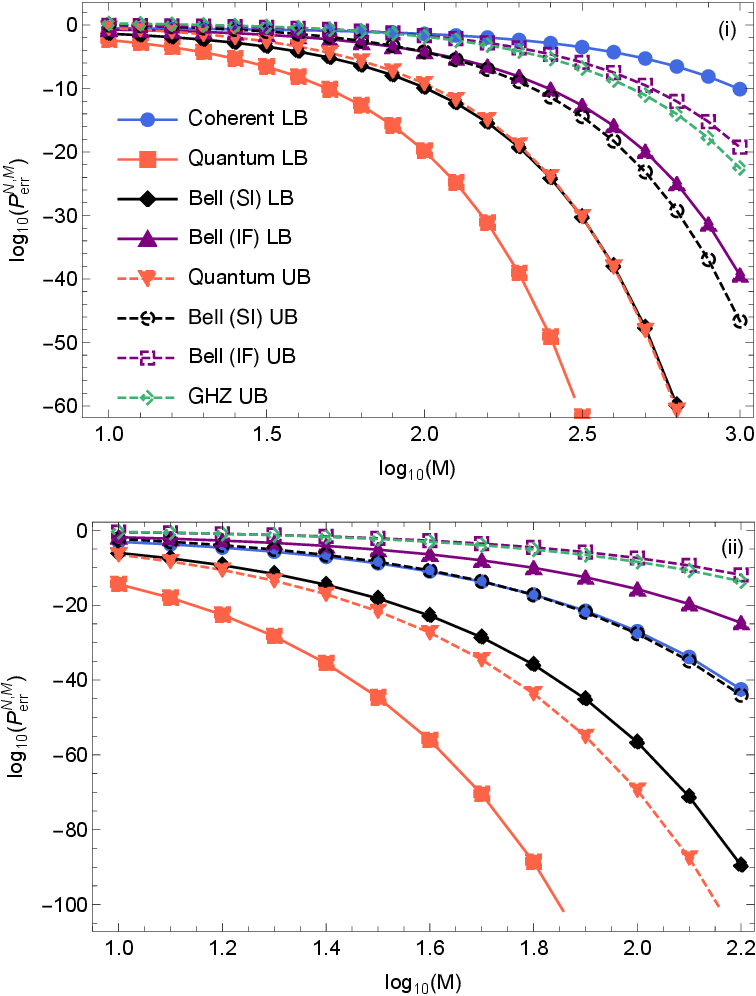}
\caption{Quantum channel position finding error probability $p_{\mathrm{err}}^{N,M}$  with $N=4$ as a function of number of uses $M$ for four types of source: 1) Coherent state (blue), 2) Quantum single photon state (red), 3) Bell biphoton state in both a signal-idler (SI) (black) and idler-free (IF) (purple) set-up, and 4) GHZ state (green). These plots show performance in locating the target channel with damping rate $\gamma_1 = 0$ (the identity channel), amongst reference channels with (i) $\gamma_0 = 0.2$, low damping, and (ii) $\gamma_0 = 0.8$, high damping. Lower and upper bounds are indicated by solid and dashed lines, respectively.}
\label{PerrBoundsIdentity}
\end{figure}

\begin{figure}[t]
\centering
\includegraphics[width=.9\linewidth]{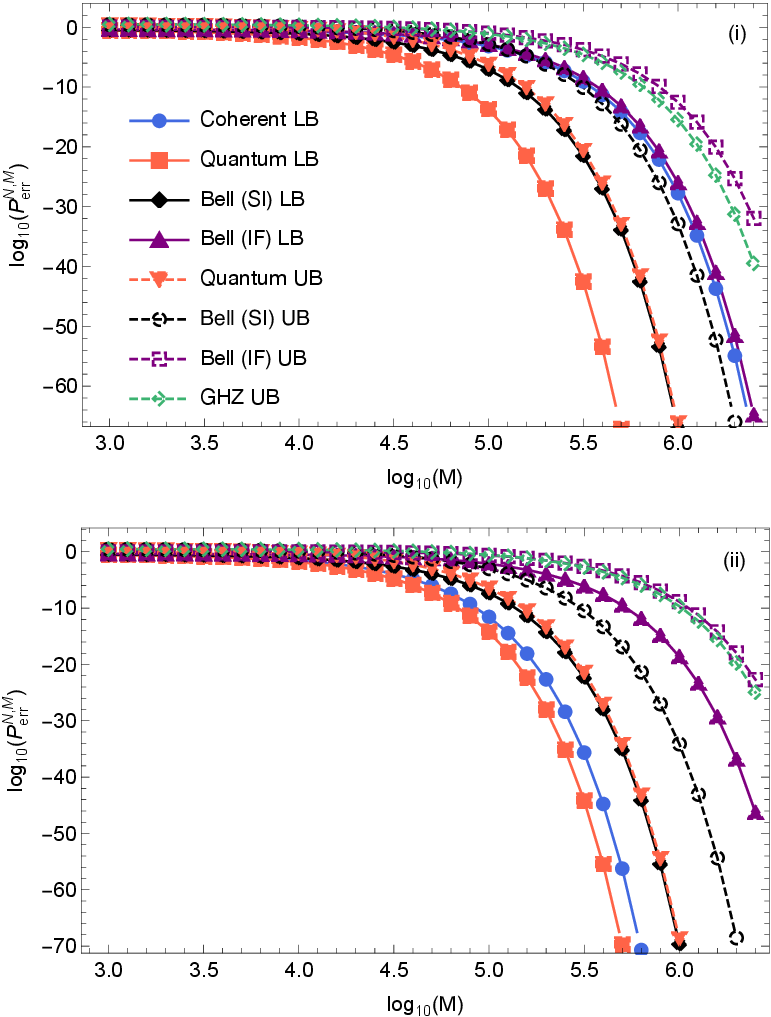} 
\caption{Quantum channel position finding error probability $p_{\mathrm{err}}^{N,M}$  with $N=4$ as a function of number of uses $M$ for four types of source:  1) Coherent state (blue), 2) Quantum single photon state (red), 3) Bell biphoton state in both a signal-idler (SI) (black) and idler-free (IF) (purple) set-up, and 4) GHZ state (green). These plots show performance in locating a single target channel with damping rate $\gamma_1$ amongst reference channels with non-zero damping rates such that $\gamma_0 = \gamma_1+0.01$ where (i) $\gamma_1 = 0.2$, low damping, and (ii) $\gamma_1 = 0.8$, high damping. Lower and upper bounds are indicated by solid and dashed lines, respectively.}
\label{PerrBoundsDelta}
\end{figure}

The task of CPF is to determine the location of the target channel, not just the pair in which it is contained. For the specific idler-free CPF protocol under consideration, there are two ways in which an overall error may be made. You may choose a pair of channels which does not contain the target then, within that pair, you always fail to identify the target. This happens with probability given by Eqs.~(\ref{idlerfreeerr1}) and~(\ref{idlerfreeerr2}). Otherwise, you successful choose the correct pair and then the task is to determine which of those two channels is, in fact, the target. To do this, since one knows the specification of the channels under study, one could engineer a secondary CPF protocol through the addition of two further reference channels on either side of the successfully located pair. Practically speaking one would simply reconsider now-known reference channels within the pattern. This would effectively realise a four-box CPF problem, in which after successfully determining the correct pair, one is certain which specific box contains the target channel. To maintain our energy constraint we choose to split our total number of probes in two, yielding $M/2$ probings for each stage of the overall idler-free CPF procedure.

Taking this two-stage approach into consideration, the $N$-box idler-free CPF protocol's error probability takes the following form, where we can employ the relevant lower and upper bounds as required:
\begin{equation}
\begin{split}
    p_{\mathrm{err}}^{N,M/2}&(\dyad{\psi_2}{\psi_2}) = \tilde{p}_{\mathrm{err}}^{N,M/2}(\dyad{\psi_2}{\psi_2}) \\
    & + \left[1- \tilde{p}_{\mathrm{err}}^{N,M/2}(\dyad{\psi_2}{\psi_2})\right] \tilde{p}_{\mathrm{err}}^{4,M/2}(\dyad{\psi_2}{\psi_2}).
    \end{split}
\end{equation}

The performance of such a biphoton state in both signal-idler and idler-free protocols is plotted in Figs.~\ref{PerrBoundsIdentity} and Fig.~\ref{PerrBoundsDelta} along with the coherent state lower bound (Eq.~(\ref{CSlowerbound})), single-photon lower (Eq.~(\ref{Focklowerbound})) and upper bound (Eq.~(\ref{Fockupperbound})), and the GHZ state's upper bound (Eq.~(\ref{GHZupperbound})). It can be seen that the biphoton Bell state's upper bound in a signal-idler protocol follows very similarly the behaviour of the GHZ state's upper bound. In general, and particularly for low damping/high transmissivity, entanglement-based protocols can yield a quantum advantage in CPF. This is especially true for idler-free protocols, as also described in Ref.~\cite{pereira2021idler} under the continuous variable formalism. While a signal-idler protocol is most advantageous, one may instead use an idler-free protocol, forgoing the need for a quantum memory, while still retaining a quantum advantage provided channel losses are not too high. Note, however, that such a scheme as described here may only be applied to CPF problems comprising an even total number of channels. While this may appear constraining, this can easily be achieved in an experiment by simply adding an extra channel when required.

\begin{figure}[t]
\centering
\includegraphics[width=.9\linewidth]{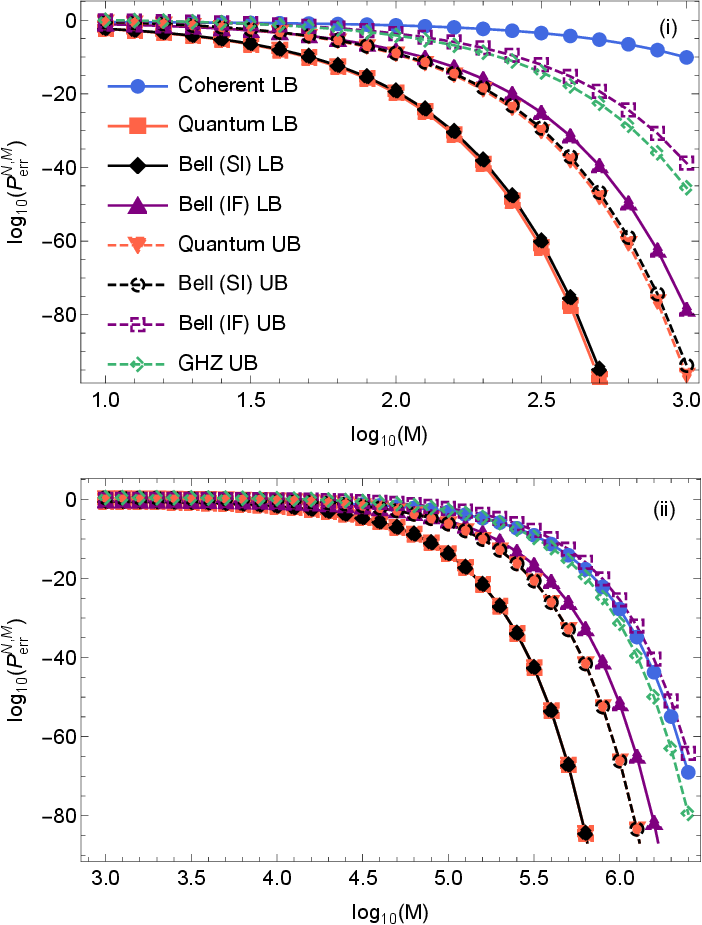}
\caption{Quantum channel position finding error probability $p_{\mathrm{err}}^{N,M}$  with $N=4$ as a function of number of uses $M$ for four types of source: 1) Coherent state (blue), 2) Quantum single photon state (red), 3) Bell biphoton state in both a signal-idler (SI) (black) and idler-free (IF) (purple) set-up, and 4) GHZ state (green). In the latter three cases, we let $M \rightarrow 2M$ to model the performances when the average number of photons used each use is equal to 1. These plots show performance in locating the target channel in the low damping regime with (i) $\gamma_0 = 0.2$ and $\gamma_1 = 0$, and (ii) $\gamma_0 = 0.21$ and $\gamma_1 = 0.2$. Lower and upper bounds are indicated by solid and dashed lines, respectively.}
\label{PerrBoundsLowEqual}
\end{figure}

Note that in all entanglement-based CPF approaches considered, namely, using the GHZ and biphoton (in both signal-idler and idler-free setups) states as sources, the average number of photons per channel use is equal to $1/2$. One can adjust the protocol across these sources to consider equal energetic distributions; letting $M \rightarrow 2M$ means that the average number of photons per channel use is equal to 1, matching that for the single photon Fock state and the coherent state classical benchmark. For completeness, Fig.~\ref{PerrBoundsLowEqual} plots the associated error probabilities for CPF arising from these changes in the low damping regime where quantum advantages are greatest, equivalent to modified versions of Fig.~\ref{PerrBoundsIdentity}(i) and Fig.~\ref{PerrBoundsDelta}(i). Both the upper and lower bounds on the error probability in the signal-idler protocol almost coincide with those of the single photon state, with those for the idler-free protocol closely following them.

\subsection{Practical receiver for quantum channel position finding based on photon counting}\label{sec:receiver}

Attainment of bounds shown in Figs.~\ref{PerrBoundsIdentity} and~\ref{PerrBoundsDelta} typically requires optimal detection at the output, the specifics of which is generally unknown. Even then they only provide bounds for the absolute performance which is given by the Helstrom limit~\cite{Helstrom,holevo1973statistical}. As a result, it is important to consider practical receiver designs to harness the input state's potential at performing the task. When the input and output states form a tensor product state across the $N$ boxes, as for the cases studied in Sections~\ref{classicalbenchmark} and~\ref{singlephoton}, we can employ a strategies based on photon-counting at the receiver end of each box and use post-processing to determine which hypothesis is true. 

Consider the use of photon counting at the receiver. Then for some generic state $\rho$ with field operator $\hat{a}$, the measurement result is a classical random variable $n$ with distribution $p(n) = \bra{n}\rho \ket{n}$, where $\ket{n}$ is the eigenstate of the number operator $\hat{n} = \hat{a}^{\dag} \hat{a}$ with eigenvalue $n$. 

The output of the $i$th box after $M$ probes is the classical string of length $M$, $\mu_i^{\otimes M}$, consisting of zeros and ones corresponding to the absence or presence of photon counts. Modelling our receiver as a threshold detector, we know that the probability of detecting $m \equiv \Tr \mu^{\otimes M}$ photons in $M$ trials follows the binomial distribution
\begin{equation}
p_{\mathrm{det}}(p,M,m) = {M \choose m} p^{m} (1-p)^{M-m},
\label{probabilityminM}
\end{equation}
with $M$ trials and probability of success $p$.

In the case of coherent state inputs, whose initial photon distribution is Poissonian, we may compute the probability of detecting $n$ photons at the output using their Fock bases such that
\begin{equation}
p_{\mathrm{C}}(n) = |\braket{n}{\tau \alpha}|^2 = e^{-\tau} \frac{\tau^n}{n!} = e^{\gamma-1} \frac{(1-\gamma)^n}{n!},
\end{equation}
where we have applied the constraint $|\alpha|=1$. Then the probability of getting a single `click' in the photon detector is
\begin{equation}
p_{\mathrm{C}}(n=1) = e^{\gamma-1}(1-\gamma).
\end{equation}

When our source consists of single qubit/photon states, then with photon counting at the output we have that 
\begin{equation}
p_{\mathrm{Q}}(n=1) = 1-\gamma,
\end{equation}
and it is clear that 

\begin{equation}
p_{\mathrm{Q}}(n=1) > p_{\mathrm{C}}(n=1) \,\, \forall \,\, \gamma \in \left[0,1\right].
\end{equation}

\begin{figure}[t]
    \centering
    \includegraphics[width=.9\linewidth]{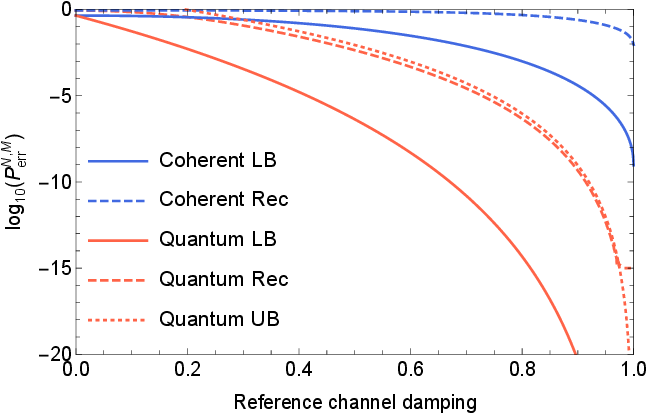}
    \caption{Quantum channel position finding error probability $p_{\mathrm{err}}^{N,M}$  with $N=10$ and $M=10$ as a function of reference channel damping $\gamma_0$ with target channel damping $\gamma_1=0$. Shown is the performance of a photon-counting receiver ``Rec" (dashed) on coherent states (blue, upper two lines) and single photon states (red, lower three lines). Lower ``LB" and upper ``UB" bounds are indicated by solid and dotted lines, respectively.}
    \label{fig:receiver_refdamp}
\end{figure}

\begin{figure}[t]
    \centering
    \includegraphics[width=.9\linewidth]{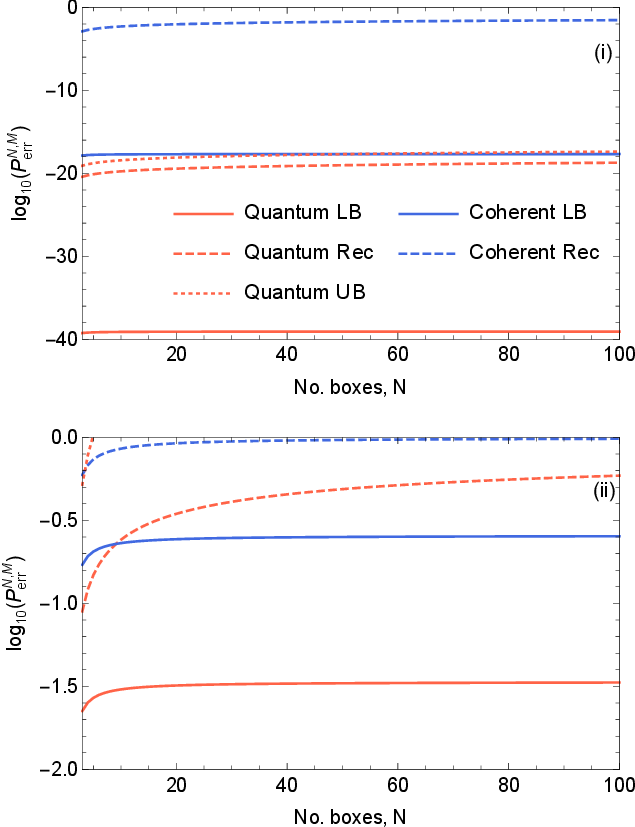}
    \caption{Quantum channel position finding error probability $p_{\mathrm{err}}^{N,M}$  with $M=100$ as a function of the total number of boxes $N$. Target channel damping $\gamma_1=0.2$ with reference channel damping set to (i) $\gamma_0=0.8$, and (ii) $\gamma_0=0.3$. Shown is the performance of a photon-counting receiver ``Rec" (dashed) on coherent states (blue, (i) upper two lines, (ii) second and fourth lines) and single photon states (red, (i) lower three lines, (ii) first, third and fifth lines). Lower ``LB" and upper ``UB" bounds are indicated by solid and dotted lines, respectively.}
    \label{fig:receiver_noboxes}
\end{figure}

Consider a relatively straightforward approach where the total number of output photons at each individual receiver are counted. Then the target channel is declared by choosing the one with the highest or lowest total photon count, dependent on the relative magnitudes of damping rates between the target and reference channels. Let the probability of measuring a photon at the output of our target(reference) channel be given by $p_T$($p_R$). Then the decision rule becomes
\begin{equation}
    \begin{cases}
    \arg \max_i m_i \quad \mathrm{if}\, p_T>p_R,\\
    \arg \min_i m_i \quad \mathrm{if}\, p_T<p_R.
    \end{cases}
\end{equation}
First consider the case where $p_T>p_R$. In such a scenario the min/max receiver would determine that the target channel is the one whose output has the highest of photon count, and would do so perfectly when all reference channel boxes output fewer photons than the target channel box. To account for the scenarios where any number of reference channels output an equal number of photons to the target channel, the decision is reduced to choosing at random between those boxes. Such a receiver design yields a total error probability 
\begin{equation}
p_{\mathrm{err}}^{\mathrm{min/max}} = 1-p_{\mathrm{succ}}^{\mathrm{min/max}},
\end{equation}
where the probability of success, $p_{\mathrm{succ}}^{\mathrm{min/max}}$, is given by
\begin{equation}
\begin{split}
p_{\mathrm{succ}}^{\mathrm{min/max}} &= \sum_{r=1}^{N}  \sum_{m_T=0}^M  p_{\mathrm{det}}(p_R,M,m_R<m_T)^{N-r} \\
&\frac{1}{r}{N-1 \choose r-1} p_{\mathrm{det}}(p_T,M, m_T) p_{\mathrm{det}}(p_R,M,m_T)^{r-1}.
\end{split}
\end{equation}
The index $r$ is the number of boxes whose output total photon counts are the same, thus requiring a random decision between them. The probability that any given reference box yields an output lower than the target channel is given by
\begin{equation}
    p_{\mathrm{det}}(p_R,M,m_R<m_T) = \sum_{m_R=0}^{m_T-1} p_{\mathrm{det}}(p_R,M,m_R),
\end{equation}
which, to allow for a random choice between all $N$ channels in the event that all outputs are 0, is defined such that
\begin{equation}
    p_{\mathrm{det}}(p_R,M,m_R<0)^0 = 1.
\end{equation}

The performance of such a photon counting receiver on CPF outputs using single photon (Fock state) sources and classical coherent state sources are shown in Figs.~\ref{fig:receiver_refdamp} and~\ref{fig:receiver_noboxes}. Fig.~\ref{fig:receiver_refdamp} shows how the error probability changes as a function of reference channel damping when the target channel is equal to the identity. Quantum-enhanced CPF using single photon states may be achieved across almost the entire range of damping values, with the simple receiver based on direct photon counting capable of realising this for most of this range. Fig.~\ref{fig:receiver_noboxes} shows how the performance of each type of source, with the use of the photon counting receiver, changes with the size of the CPF array. This behaviour is shown for the addition of background channels of differing damping strength relative to the target channel of interest. It can be seen that, in general, coherent state CPF capabilities are more robust to increasing pattern size in that the performance remains fairly constant compared to Fock states with added reference channels. As one would expect, both classical and quantum sources' performances are affected more detrimentally when the background channels added are relatively similar in specification to the target channel, though this effect is greater for single photon states employed in conjunction with our proposed photon counting receiver.

\section{Conclusion}
This paper provides performance comparisons of various types of source with respect to the problem of CPF where the channels are of differing transmissivity. The work is carried out in the discrete-variable setting under the energetic constraint that each channel use (probing of the entire channel array) is limited to use, at most, one photon on average.

Considered in this work are the performances of various types of discrete variable quantum sources, namely, the single-photon (Fock) state, the GHZ state where entanglement is evenly distributed across the CPF channel array and the biphoton (Bell) state.  In the latter case, two protocols are considered: one considering signal-idler (QI) role across the two modes and another where both modes are used as a signal. In all cases, fidelity-based bounds on error probability are derived.

While the signal-idler approach for biphoton states tends to be sub-optimal compared to the use of Fock states, it is the most optimal amongst the entanglement-based approaches considered. Nonetheless, such a method may be experimentally demanding since it requires access to a quantum memory. Our idler-free protocol forgoes this need and still retains much of the desired quantum advantage in cases where channel damping is low. The biphoton state is one which may be readily generated and employed using MRRs which could form the basis of a proof-of-principle experiment for quantum-enhanced CPF using single photon states.

\textbf{Acknowledgments.}~This work has been funded by the European Union's Horizon
2020 Research and Innovation Action under grant
agreement No. 862644 (FET-Open project: Quantum readout techniques
and technologies, QUARTET). AK acknowledges sponsorship by EPSRC Award No. 1949572 and Leonardo UK. SP would like to thank enlightening discussions with Janos a Bergou and Mark Hillery.

\end{document}